# Mapping EINS

## An exercise in mapping the Network of Excellence in Internet Science

Almila Akdag Salah
New Media Studies
University of Amsterdam
Amsterdam, The Netherlands
A.A.Akdag@uva.nl

Sally Wyatt / Samir Passi / Andrea Scharnhorst
eHumanities group
Royal Netherlands Academy of Arts and Sciences
Amsterdam, The Netherlands
Andrea.scharnhorst@dans.knaw.nl

***Abstract*—This paper demonstrates the application of bibliometric mapping techniques in the area of funded research networks. We discuss how science maps can be used to facilitate communication inside newly formed communities, but also to account for their activities to funding agencies. We present the mapping of EINS as case – an FP7-funded Network of Excellence. Finally, we discuss how these techniques can be used to serve as knowledge maps for interdisciplinary working experts.**

### I. Introduction

The academic world of science and technology represents an ever-changing landscape characterized by the continuous emergence and development of new research directions, funding initiatives, scientific publications, and communication and collaboration networks. One of the ways to represent and analyze this dynamic landscape is through science maps. Science maps are two- or three-dimensional visualizations that represent specific attributes pertaining to particular scientific field(s). These attributes can be related to research themes within disciplines, results of scientific experiments, or the nature of citations within and between disciplines. Examples of science maps include, but are not limited to, maps representing a) bibliometric analysis of scientific publications, b) collaborative research networks, c) nature and quantity of research funding, and d) the trajectory of the evolution of scientific fields. In this paper, we are specifically interested in science maps representing bibliometric analysis of research publications.

Scientific and research publications provide a relevant point of entry to study the nature, trajectory, and structure of scientific fields. [9] Bibliometric records contain a wide array of information such as author name(s), title, year of publication, citations, publication type etc. Each of these elements, or combinations thereof, can be employed to generate unique science maps representing different aspects of the relationships between and within scientific disciplines and research communities. For example, a science map based on publication types (differentiating between types of journals for example) helps showcase the dominant disciplinary and research fields while a map based on co-occurrence of authors (two or more authors co-authoring a publication) represents the formulation and evolution of collaborative networks between disciplines and research communities. [6] Science maps, thus, are useful tools to understand the state-of-the-art and disciplinary structure within an academic field as well as to analyze the emergence of research networks and collaborations. Such an understanding can help provide relevant insights not only into which specific disciplinary areas need more focus and research efforts but also into the ways in which collaborative research networks can be sustained and developed further.

The very first attempts to map science [5] were meant to visualize research fronts and most intriguing new areas in the sciences [13]. Current science maps displayed the structure of academia mining and aggregating millions of links in scholarly communication (such as co-words, citations, co-citations) in a bottom-up approach to identify higher level structures in science such as disciplines. [1] Soon these so-called backbone maps of science were used as reference systems against which the position of a lab, an institute or a whole country could be "science-located". In this function they can be used as a heuristic for studying the sciences; but also as a means for evaluation and science policy advice. Currently, there are a number of tools available, such as Sci2Tool [12], CiteSpace [2], Vosviewer [14] and OverlayToolkit [10]. But, most of the analysis has been done *about* communities, programmes, research trends. The argument we bring forward in this paper is to use science mapping as internal communication tool *by* the communities themselves.

### II. The use case – EINS

EINS - **E**xcellence in **IN**ternet **S**cience[1] - is a so-called Network of Excellence (NoE) project funded under the 7$^{th}$ Framework programme. This funding scheme is meant to "contribute to the clarification of the concepts in the covered field"[2] and to coordinate research capacities, rather than directly to fund research. Network activities such as workshops and research mobility are therefore central for this project. EINS started on 1 December 2011 and continues for 42 months. About 5 million euros have been allocated to a consortium formed by 33 research institutes from 16 countries

[1] www.internetscience.eu

[2] http://en.wikipedia.org/wiki/Framework_Programmes_for_Research_and_Technological_Development

(including partners from China, South Korea, Australia and Canada). The main goal is to achieve "a deeper multidisciplinary understanding of the development of the Internet as a societal and technological artifact". [16]

From the kick-off plenary meeting onwards is was obvious that experts in the project come from very different scientific disciplines, and that to exchange, communicate and negotiate notions, concepts, and methods around a phenomena as broad as the Internet will become a real challenge. [3] This observation prompted this study in bibliometric mapping of the community.

The research strategy was to "science-locate" the current position of the network of excellence by looking into their publications prior to EINS, to detect the knowledge base of community in terms of most cited literature, and to analyze the co-authorship network. To compare the EINS network with current other research about the Internet, we did a topical search for the term "Internet" and the year 2012 in the Web of Science bibliographic database. Publications from this sample are used to contrast the occupation of the EINS network on a global map of science with an occupation of Internet-related publications on the same map.

## III. Data collection and applied tools

### A. Data

The data source for this analysis was the Web of Science™ (WoS) – an interdisciplinary and international bibliographic database, which includes mostly journal publications, and with them also the cited references. [4] In October 2012 we retrieved a set of 1353 articles by searching for the names of 118 EINS members, which were listed on the EINS website at that time. We searched for their publications using the author search function. Few of the experts had no publications in the WoS. For most of them the search delivered different listings (pre-aggregated author sets). Those could be sorted out using information about the academic background and therefore the publications venues where we expect them to publish, and/or using institutional information (for the problem of author ambiguity see [11]). One Chinese name could not be retrieved properly, and was excluded from the analysis eventually. After retrieving all articles we used the Sci2Tool to limit the publication years to the five years prior to the start of the EINS.

We collected a second dataset using the topical search function in WoS and "Internet" as search term. The topical search function retrieves publications which contain the search term either in the "Title", "Abstract", "Author Keywords" or "Keywords Plus®" field of the bibliographic record. We restricted our search to the year 2012 only. The search was performed in December 2012 and delivered 1191 records. We would like to emphasise that such a data mining strategy is expected to give only a global and rough impression of the activities related to Internet Science. The internet is not that new any more so authors may no longer feel justified in putting "Internet" in the title of each work which is relevant for the study of the Internet. Consequently we use this dataset as an illustration for the expansion of 'internet' literature across a global map of science, than for an analysis of current Internet Science. The latter would require a much more specifically designed and larger data mining strategy. Both datasets are available from the authors for inspection.

### B. Tools

We used the Sci2 Tool of the Cyberinfrastructure for Network Science Center, Indiana University Bloomington for data cleaning, pre-analysis, and the co-authorship network. We used network mapping and analysis tools (including the Overlay toolkit) provided by Loet Leydesdorff for the other maps[3]. We used *Gephi* for the final network layout, and *Adobe Illustrator* for the final design.

## IV. Results

In this section we present the maps and discuss them.

### A. Distribution of publications of EINS' experts

This map (Fig. 1) displays publications of the EINS community over a general map of science (first dataset of 1353 records). The transparent network structure of the general map of science is colour-coded according to different disciplines. Using the overlay technique developed by Ismael Rafols and others [10], areas in which members publish appear as nodes, colour-coded according to the assigned discipline and size coded.

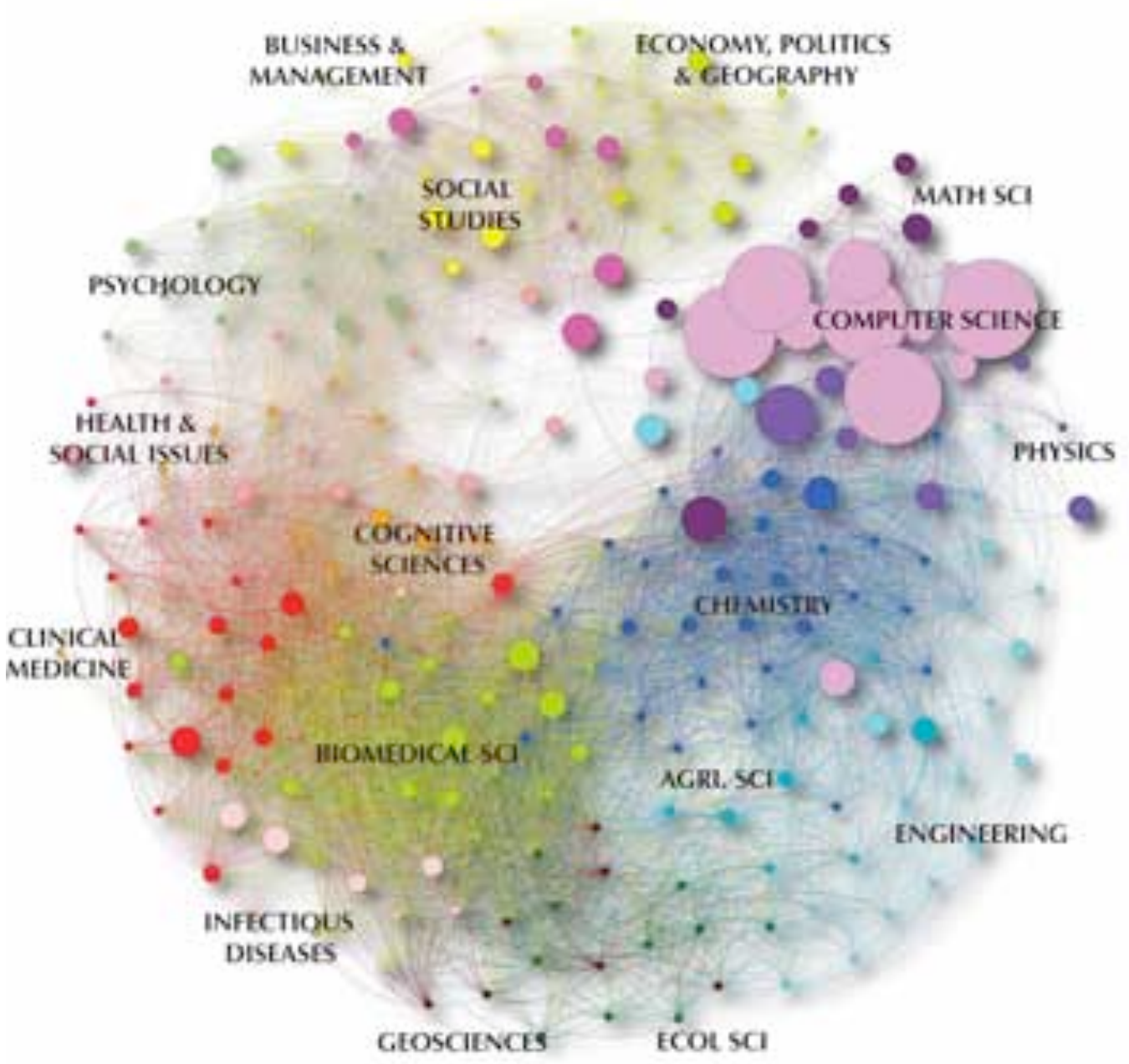

Fig. 1. Publications of EINS experts on the global map of science

The visualization (Fig. 1) shows the main output disciplines of the members: computer science and physics, 1303 of the total of 1353 articles belong to those disciplines. Using an absolute scale of node size coding made the already existing outreach into other areas invisible. We therefore used the spline-size coding option for node size scales of the visualization software *Gephi*.

[3] http://www.leydesdorff.net/software.htm

## *B. The knowledge base of EINS*

This map confirms the dominance of computer and natural sciences in this expert network and at the same time indicates the challenge for this community. The map uses the set of 1353 articles and displays the network of cited journals – the knowledge base of this community. Whenever two journals appear in the reference list of an article a link is established between them. Apart from the interdisciplinary journals *Science* and *Nature*, the journal landscape of this community, and with it their conceptual frameworks, methods and use of language is dominated by computer sciences, physics and engineering.

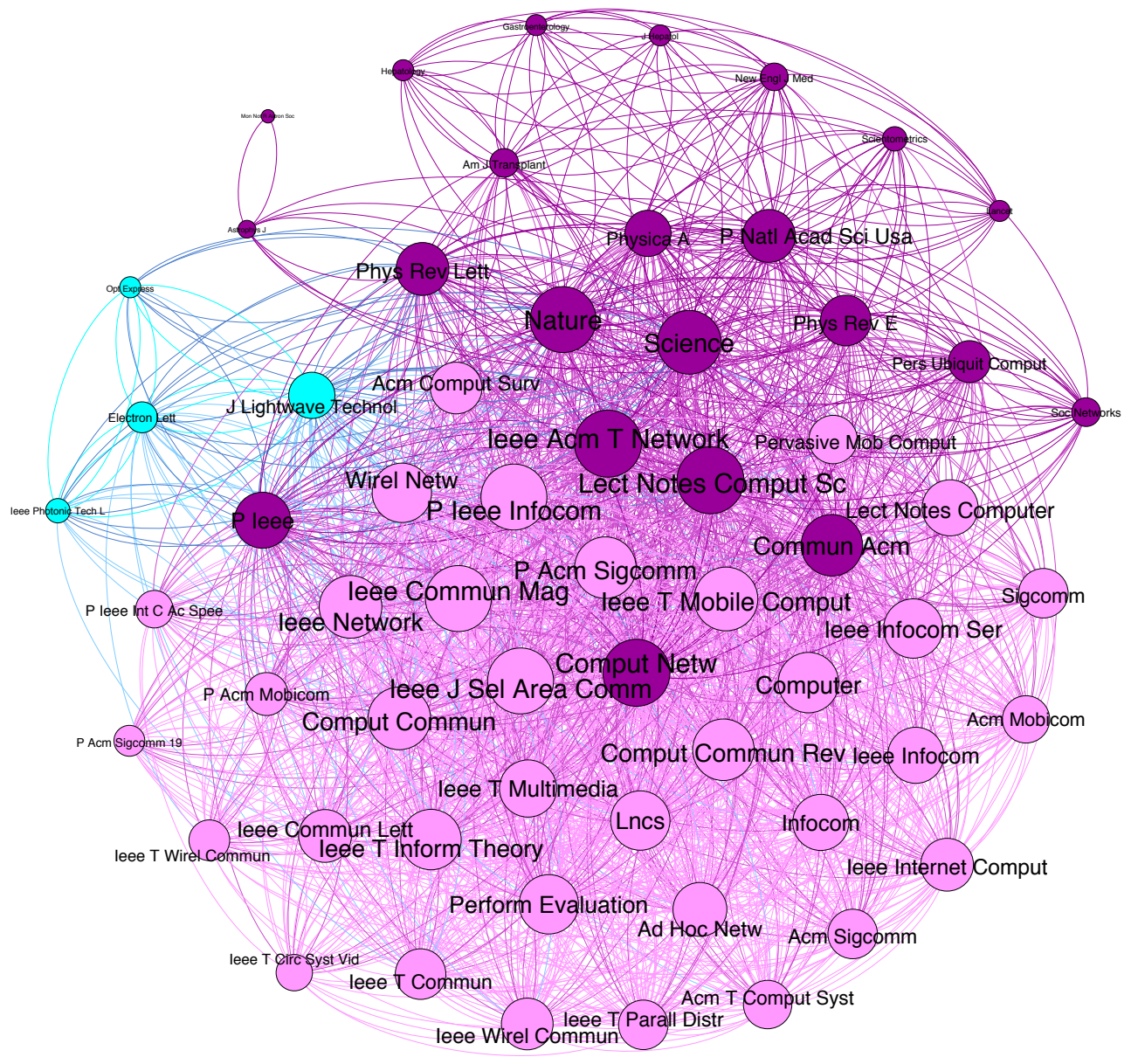

Fig. 2. Network of journals relevant for the EINS community

## *C. Co-author map of EINS*

The map shows the existing co-authorship network based on the same data set of 1353 articles of EINS members. A program (co-author.exe) devised by Loet Leydesdorff has been used for this analysis. The resulting network (threshold value 0.1 %) shows 122 connected authors, 57 of whom are members of EINS. Only the EINS member nodes are coloured (according to the subject categories in which they actively publish), the rest of the nodes are marked with grey. The node sizes correspond to the in-degree parameter, i.e. the more co-authors an author has, the bigger the node size. Naturally EINS members come with their existing collaboration networks. One goal of this Network of Excellence is to sustain and widen those collaborative networks.

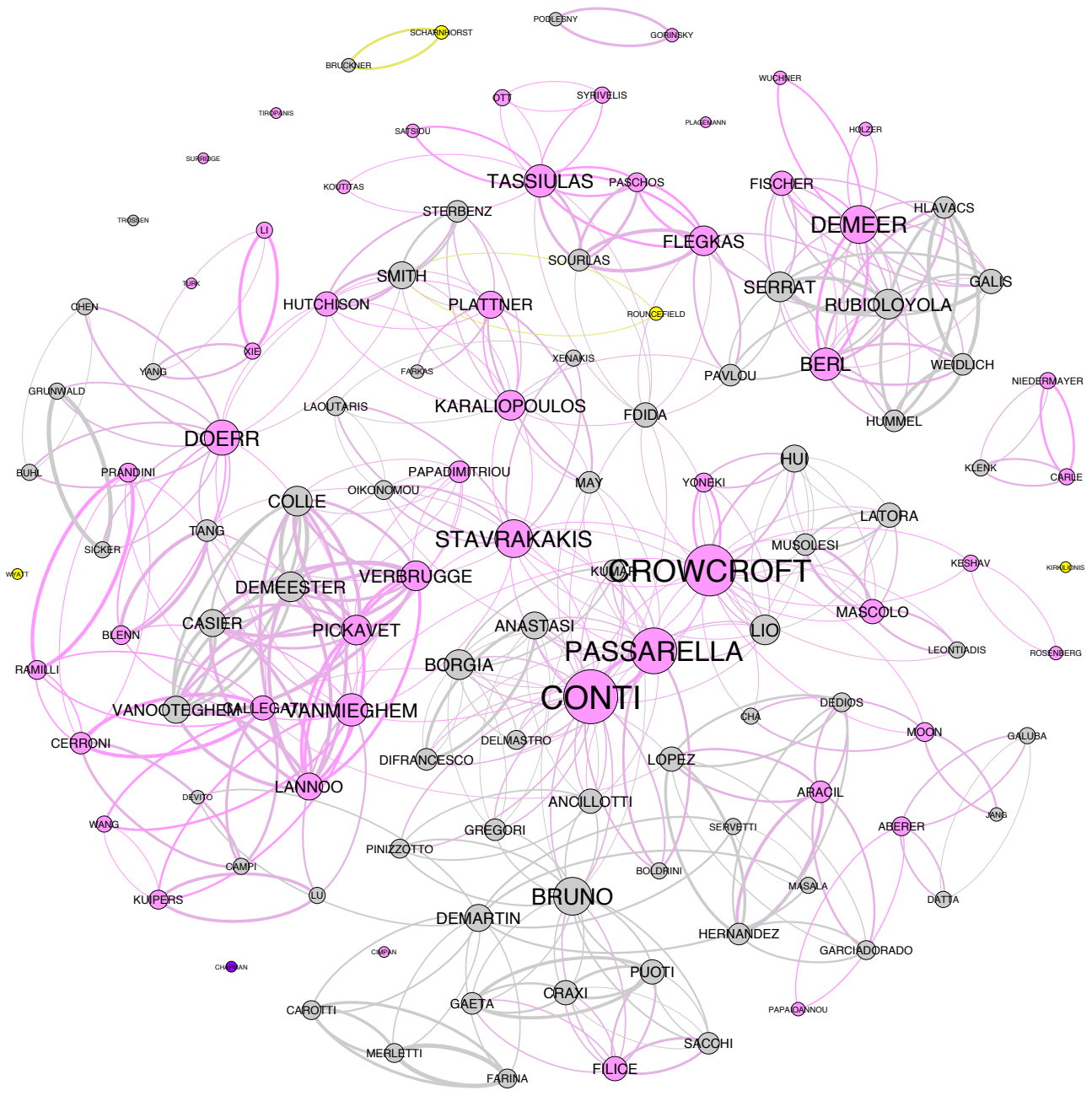

Fig. 3. Existing collaboration network at the start of EINS

## *D. Overview map of 'Internet Science'*

The current world of 'internet science' or 'internet research' encompasses all scientific disciplines. This map shows the overlay map of our second dataset (1191 records), based on a search of the term "Internet" in titles, etc. of articles in the Web of Science for the year 2012 only. Using the same overlay technique as in Fig. 1 this visualization shows that computer science and physics are not the only leading research areas to study the Internet. From anthropology, psychology, economy, politics to medical and cognitive research and public health issues, almost all areas of the science system are concerned with the Internet. One goal of the EINS community is to link its core competences with other Internet Science research threads, to occupy more areas in the overall 'internet science landscape', and to contribute to a consolidation of the knowledge about the Internet.

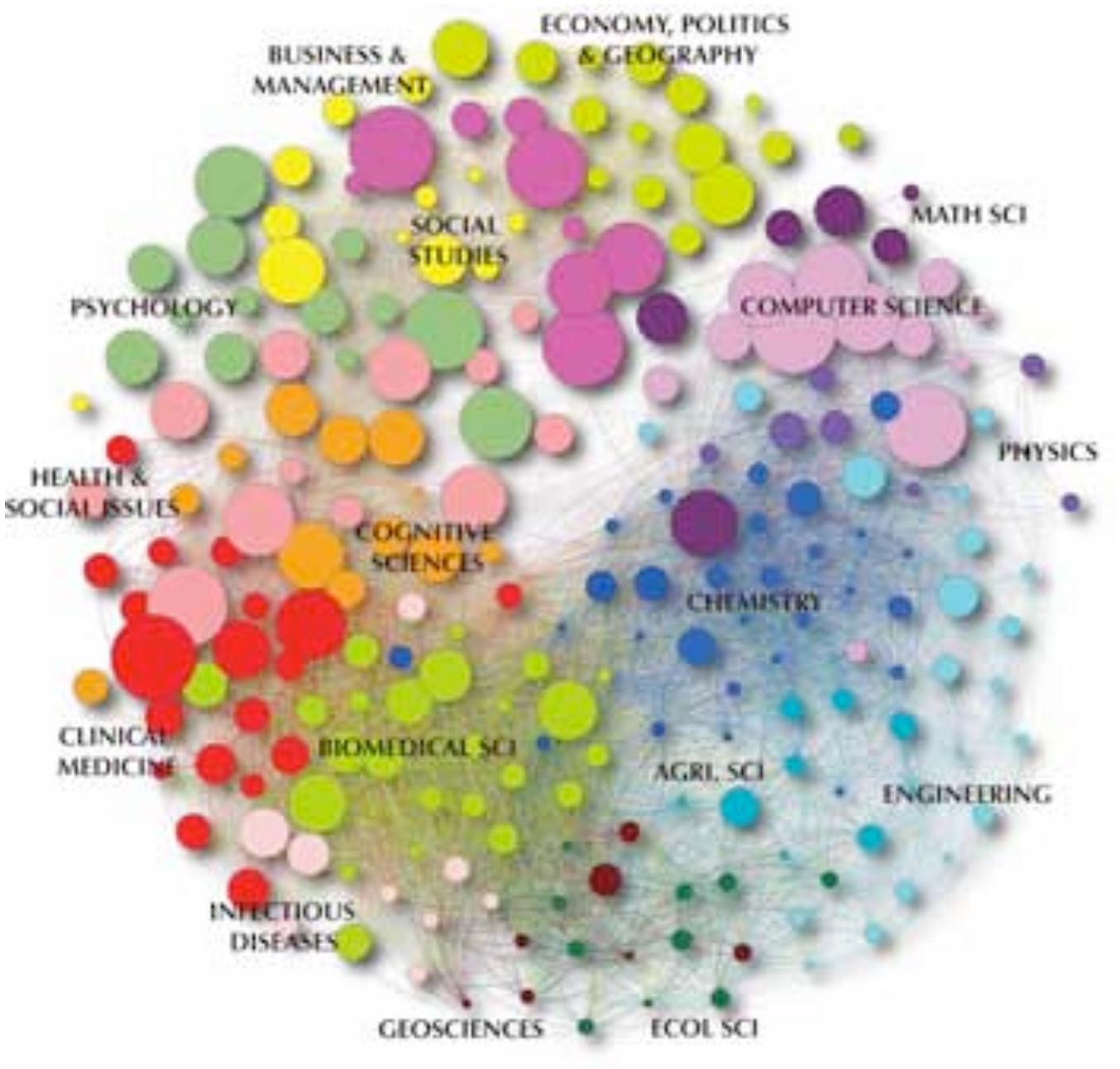

Fig. 4. Internet-related publications at the global map of science in 2012.

## V. Conclusions

The exercise described above has implications not only for the EINS community, but also for research agenda setting and evaluation. First, for the EINS community, the results indicate that there is an imbalance between the disciplines predominant in EINS and the full range of scientific and scholarly work being conducted about the 'Internet'. Given the massive diffusion and use of the Internet since it went public and commercial in the mid-1990s, it is not surprising that much academic attention about its effects has come from the full range of disciplines. Understanding the Internet requires input from all. Thus, EINS is correct in its policy of attempting to develop the internet research community and to build bridges between disciplines through its various networking activities. Second, such maps can be used to identify lacunae for further research. However, this would need to be combined with more qualitative approaches. Recently, van Heur, Leydesdorff and Wyatt [8] did something similar, combining the kinds of analyses described above with close reading of texts in order to assess the extent and nature of the ontological turn in science and technology studies. Third, such maps can be used as evaluation tools. After EINS is finished, a similar exercise can be conducted in order to assess whether EINS was successful in promoting interdisciplinary collaboration by examining co-authorship and co-citation patterns in future work. [7] Fourth, science mapping techniques combined with automatic information retrieval allow for an fast and easy overview about future emerging trends, under-occupied research areas and missing links in research collaboration. In this capacity, science maps or more general knowledge maps have potential as visual enhanced navigation tools in a large and complex spaces of information [15].

## Acknowledgment

This work has been funded by the Network of Excellence for Internet Science, FP7 – 288021.